\documentclass[12pt]{elsarticle}

\usepackage{amssymb}
\usepackage{amsmath}
\usepackage{graphics}
\usepackage{latexsym}
\usepackage{array}
\usepackage[activeacute,english]{babel}
\usepackage[latin1]{inputenc}

\usepackage{color}
\usepackage{lineno}
\usepackage{latexsym}
\usepackage[colorlinks=true]{hyperref}

\newcommand{\bn}{\begin{eqnarray}}
\newcommand{\en}{\end{eqnarray}}
\newcommand{\bq}{\begin{equation}}
\newcommand{\eq}{\end{equation}}

\newcommand{\bc}{\begin{center}}
\newcommand{\ec}{\end{center}}

\journal{Annals of Physics}

\begin{document}

\begin{frontmatter}

\title{Dirac equation in low dimensions: The factorization method}

\author[usp]{J. A. S\'anchez-Monroy\corref{cor1}}
\ead{antosan@if.usp.br}
\author[rvt,focal]{C. J. Quimbay\corref{cor2}}
\ead{cjquimbayh@unal.edu.co}

\address[usp]{Instituto de F\'\i sica, Universidade de S\~{a}o Paulo, 05508-090, São Paulo, SP, Brazil}
\address[rvt]{Departamento de F\'\i sica, Universidad Nacional de Colombia,
Bogot\'a, D. C., Colombia}
\address[focal]{Associate researcher of CIF, Bogot\'a, Colombia}
\cortext[cor1]{Corresponding author}
\cortext[cor2]{Ciudad Universitaria, Building 404, Room 343, Bogot\'a,
Colombia. Phone: (57)(1)3165000 Ext. 13051.}

\begin{abstract}
We present a general approach to solve the $(1+1)$ and
$(2+1)$-dimensional Dirac equations in the presence of static
scalar, pseudoscalar and gauge potentials, for the case in which
the potentials have the same functional form and thus the
factorization method can be applied. We show that the presence of
electric potentials in the Dirac equation leads to two
Klein-Gordon equations including an energy-dependent potential. We
then generalize the factorization method for the case of
energy-dependent Hamiltonians. Additionally, the shape invariance
is generalized for a specific class of energy-dependent
Hamiltonians. We also present a condition for the absence of the
Klein paradox (stability of the Dirac sea), showing how Dirac
particles in low dimensions can be confined for a wide family of
potentials.
\end{abstract}

\end{frontmatter}

\newpage
\section{Introduction} \label{Sec1}

The Dirac equation in the presence of different types of
potentials leads to the prediction of some peculiar effects.
Specifically, the presence of an electric scalar potential in the
Dirac equation can imply the existence of the Klein paradox
(instability of Dirac sea). However, in the context of elementary
particle physics, this effect has not been experimentally observed
until now due to the impossibility of having a large enough
electric field. Thus, the effects produced by different kind of
potentials in the Dirac equation can be studied by using quantum
simulators \cite{Gerri1,Gerri2,Casanova}. For instance, Casanova
{\it et al.} have shown how to engineer scalar, pseudoscalar and
other linear potentials in the $(1+1)$-dimensional Dirac equation
by manipulating two trapped ions \cite{Casanova}. Additionally, it
is well known that the $(2+1)$-dimensional Dirac equation
describes the low energy excitations of graphene
\cite{Castro2006}. In this sense, graphene might be considered as
a simulator of quantum relativistic effects in $(2+1)$ dimensions.
In particular, graphene allows a close realization of the Klein
\textit{gedanken} experiment \cite{Kats} and this fact might be fundamental
in the future design of graphene-based electronic devices.
\par
Solutions of the Dirac equation in the presence of diverse types
of potentials have been extensively addressed
\cite{BagGit}-\cite{Beckers}. Among the different approaches to solve
the Schrödinger and Dirac equations, we can find the
factorization method, which is based on fundamental ideas proposed
by Dirac \cite{bookdirac} and Schrödinger \cite{Schro1}. This
approach was subsequently generalized in a systematic way by
Infeld {\it et al.} \cite{Infeld}. However, it is important to
mention that the factorization method works in an equivalent way
as the Darboux transformation does \cite{Darboux1}. Additionally,
the factorization method is analogous to the so called
supersymmetric quantum mechanics (SUSYQM) proposed by Witten
\cite{Witten}. The SUSYQM has been applied to solve exactly
problems in non-relativistic and relativistic quantum mechanics
\cite{Coo}. The SUSYQM technique allows to construct a pair of
isospectral Hamiltonians (the so called supersymmetric partners) in
terms of the ladder operators, which can be used to obtain the
eigenvalues and eigenfunctions of the problem. In 1983, Gendenshtein
\cite{Gende} showed that if the supersymmetric partners are invariant under a discrete
reparametrization (usually called shape invariance), the spectrum
can be obtained algebraically.
\par

The main goal of this work is to present a general approach to
solve the $(1+1)$ and $(2+1)$-dimensional Dirac equations in the
presence of static scalar, pseudoscalar and gauge potentials using
the factorization method. We implement this method for the case in
which the potentials in the Dirac equation have the
same functional form and only depend on one
spatial cartesian coordinate. For the case in which the Dirac
equation involves the presence of an electric scalar potential, we
show that the two Klein-Gordon equations associated to the Dirac
equation include an energy-dependent potential. Due to the fact that
the factorization method has been only implemented, until now, for the
case of energy-independent Hamiltonian, we generalize the
factorization method for the case of energy-dependent
Hamiltonians. Additionally, we generalize the shape invariance for
a restricted class of energy-dependent Hamiltonians. Finally, we
present a condition for the absence of the Klein paradox and
therefore for the possible confinement of fermions. This condition
is closely related to the existence of supersymmetric partners in
the problem and this fact ensures that the Dirac sea is stable and
does not mix the positive and negative energy states, as it was
noted before by Martinez {\it et al.} in $(3+1)$-dimensions
\cite{MartinezR}. The hidden supersymmetry of the $(3+1)$-dimensional
Dirac oscillator has been discussed by Benitez {\it et al.} \cite{Benitez}
and the implication of supersymmetry on the stability of the Dirac sea
was noted. Supersymmetry in the context of the Dirac equation has also been studied by
other authors \cite{Hughes,Beckers}.

The structure of this paper is the following: First, in section
\ref{Sec1}, we present a general method to solve the
$(1+1)$-dimensional Dirac equation in the presence of static scalar,
pseudoscalar and gauge potentials, for the case in which these
potentials have the same functional form and thus the
factorization method can be applied; in section \ref{Sec2}, we
first generalize the factorization method for the case of
energy-dependent Hamiltonians, then we extend the shape
invariance for a restricted class of energy-dependent Hamiltonians
and finally we present some examples; in section \ref{Sec3}, we
show how this method can be applied to the case of the
$(2+1)$-dimensional Dirac equation, but considering static external
potentials depend only on one cartesian
coordinate; in section \ref{Sec4}, we present a condition for the
absence of the Klein paradox and therefore for the possible
confinement of fermions; finally in section \ref{Sec5}, we present the
conclusions of this work.

\section{The (1+1)-dimensional Dirac equation} \label{Sec2}

\subsection{Dirac equation in the presence of static potentials}\label{Mispv}

The (1+1)-dimensional Dirac equation in the presence of static
scalar $V(x)$, pseudoscalar $P(x)$ and gauge
$A_{\mu}(x)=\{A_t(x),A_x(x)\}$ potentials is written as
\begin{equation}\label{ED}
\left[i\gamma^{\mu}(\hbar\partial_{\mu}+i\frac{e}{c}A_{\mu}(x))-\frac{1}{c}P(x)
\gamma^{5}-\frac{1}{c}V(x)-mc\right]\Psi(x,t)=0,
\end{equation}
where the Dirac matrices $\gamma^{\mu}$, with $\mu=0,1$, are the
generators of the two-dimensional flat space-time Clifford algebra
given by
\begin{equation}
\{\gamma^{\mu},\gamma^{\nu}\}=2\eta^{\mu \nu},
\end{equation}
and $\eta^{\mu \nu}=diag(+,-)$. We choose the Dirac matrices to
satisfy the following properties
\begin{equation}\label{gammamu}
(\gamma^0)^{\dag}=\gamma^0 \ \ \text{and} \ \ (\gamma^1)^{\dag}=-\gamma^1.
\end{equation}
The Dirac equation (\ref{ED}) can be rewritten as follows ($\beta=\gamma^{0}$,
$\alpha=\gamma^{0}\gamma^{1}$)
\begin{equation}
i\hbar\frac{\partial}{\partial
t}\Psi(x,t)=\mathcal{H}\Psi(x,t)=\left(c\alpha p+ \beta
mc^2+\mathcal{V}\right)\Psi(x,t), \label{DE2}
\end{equation}
where
\begin{equation}
\mathcal{V}=\mathbf{1} eA_t(x)-\alpha eA_x(x)+\beta V(x) +\beta
\gamma^5P(x).
\end{equation}
In the last expression, $\mathbf{1}$ stands for the $2×2$ identity
matrix and $\mathcal{H}$ is the Hamiltonian operator. Because there
are only four linearly independent $2×2$ matrices, then the
potential matrix $\mathcal{V}$ is the most general combination of
Lorentz structures \cite{Castro4, Castro2003a,Castro2003b}.
\par
Let us consider the adjoint Hamiltonian $(\mathcal{H}^{\dag})$
given by
\begin{equation}
\mathcal{H}^{\dag}=\left(c\alpha p+\beta mc^2+\mathbf{1} eA_t(x)-\alpha eA_x(x)+
\beta V(x) +(\gamma^5)^{\dag} \beta (P(x))^*\right)
\end{equation}
where $^*$ denotes the complex conjugate. It was supposed that
$A_t(x)$, $A_x(x)$ and $V(x)$ are real-valued functions and we use
the properties (\ref{gammamu}). Since $\mathcal{H}$ must be a
Hermitian operator, there are two options:
$(\gamma^5)^{\dag}=-\gamma^5$ and $P(x)^*=P(x)$ or
$(\gamma^5)^{\dag}=\gamma^5$ and $P(x)^*=-P(x)$. In the next, we
use the first option and then all the potentials are real
functions. The ($1+1$) dimensional stationary Dirac equation with
a non-Hermitian and pseudoscalar interaction, has been examined
\cite{Sinha1,Sinha2,Santos1}. In the next, we will focus on
in Hermitian potentials.
\par
In order to use an appropriate representation of the Dirac
matrices $\gamma^{\nu}$, we choose
\begin{equation}
\gamma^{0}=\sigma^1=\left( \begin{array}{cc}
0 & 1 \\
1 & 0 \\
\end{array} \right), \ \ \ \ \  \gamma^{1}=i\sigma^3=\left( \begin{array}{cc}
i & 0 \\
0 & -i \\
\end{array} \right),
\end{equation}
\begin{eqnarray}
\gamma^{5}=i
\gamma^{0}\gamma^{1}=i\sigma^2=\left( \begin{array}{cc}
0 & 1 \\
-1 & 0 \\
\end{array} \right).
\end{eqnarray}
We can write the spinor $\Psi(x,t)$ as
\begin{equation}
\Psi(x,t)=e^{-\frac{iEt}{\hbar}}\left( \begin{array}{c}
\psi_1(x)  \\
\psi_2(x)  \\
\end{array} \right),
\end{equation}
and then, from the Dirac equation (\ref{DE2}), we obtain the
following coupled equation system
\begin{eqnarray}\nonumber
-c\hbar\frac{d}{dx}\left( \begin{array}{c}
\psi_1(x)  \\
\psi_2(x)  \\
\end{array} \right)+\left( \begin{array}{cc}
ieA_x(x)-V(x) & -eA_t(x)-P(x)\\
eA_t(x)-P(x)  & ieA_x(x)+V(x) \\
\end{array} \right)\left( \begin{array}{c}
\psi_1(x)  \\
\psi_2(x)  \\
\end{array} \right)&&\\\label{DEUNOMASUNO}
=\left( \begin{array}{cc}
mc^2 & -E \\
E & -mc^2 \\
\end{array} \right)\left( \begin{array}{c}
\psi_1(x)  \\
\psi_2(x)  \\
\end{array} \right).&&\label{EDCS}
\end{eqnarray}
In the next, we will not consider the spacial component of the
gauge potential $A_x(x)$ because the Coulomb gauge ($\nabla \cdot
\vec{A}=0$) can be used. For this reason, the condition
$\frac{\partial A_x}{\partial x}=0$ is satisfied for the case of a
$(1+1)$-dimensional static field. In general, this term can be
canceled by means of a gauge transformation \cite{Castro2003b}.
Thus, the problem of the Dirac equation in the presence
of a gauge potential $A_{\mu}(x)$ can be studied consistently by
considering only the presence of a static electric scalar
potential $A_t(x)$ in the system.

\subsection{Implementation of the factorization method}\label{Mispv}

Now we implement the factorization method to obtain the spectrum
associated to the Eq. (\ref{EDCS}). To do it, we restrict the
treatment to the case of static potentials with the same
functional form. In other words, these potentials only may
differ from each other by a proportionality factor or a shift in
their origin. For this reason, we assume that the
potentials have the following form $V(x)=\zeta_1
f(x)+\tilde{m}c^2$, $P(x)=\zeta_2 f(x)+\varepsilon$ and
$eA_t(x)=\zeta_3 f(x)+\tilde{E}$, where $f(x)$ is a function on
the variable $x$, and $\zeta_i$, $\varepsilon$, $\tilde{m}c^2$ and
$\tilde{E}$ are real parameters. For this kind of potentials, the
coupled equation system (\ref{EDCS}) can be written as
\begin{eqnarray} \nonumber
&&-c\hbar\frac{d}{dx}\left( \begin{array}{c}
\psi_1(x)  \\
\psi_2(x)  \\
\end{array} \right)+V(x)\left( \begin{array}{cc}
-1 & -\ell-\zeta\\
\ell-\zeta & 1 \\
\end{array} \right)\left( \begin{array}{c}
\psi_1(x)  \\
\psi_2(x)  \\
\end{array} \right)\\ \label{DEEMESD}
&&=\left( \begin{array}{cc}
(m+\tilde{m})c^2 & -(E-\tilde{E})+\varepsilon \\
(E-\tilde{E})+\varepsilon & -(m+\tilde{m})c^2 \\
\end{array} \right)\left( \begin{array}{c}
\psi_1(x)  \\
\psi_2(x)  \\
\end{array} \right),\label{EDSV1}
\end{eqnarray}
where $\zeta=\zeta_2/\zeta_1$ and $\ell=\zeta_3/\zeta_1$. We
observe that if $\zeta=0$, then $\zeta_2=0$ and there is
no presence of a pseudoscalar potential. Additionally, for
$\ell=0$ we have $\zeta_3=0$, then for this case there is no presence
of an electric scalar potential in the system. Without
loss of generality, we set $\tilde{m}=\tilde{E}=0$, because
$\tilde{m}$ and $\tilde{E}$ represent only a shift for the mass
and energy terms. Now, for this kind of potentials, we can
implement the factorization method. To implement this method, it
is first necessary to diagonalize the matrix, which is multiplying
$V(x)$, by multiplying it by a matrix $D$ from the left and by
$D^{-1}$ from the right. The matrix $D$ is given by
\begin{equation}
D=\frac{1}{2\tau(1+\tau)}\left( \begin{array}{cc}
1+\tau & \zeta+\ell \\
\ell-\zeta  & 1+\tau \\
\end{array} \right),
\end{equation}
where $\tau=\sqrt{1+\zeta^2-\ell^2}$. Introducing the notation
\begin{equation}\label{NODE}
D\left( \begin{array}{c}
\psi_1(x)  \\
\psi_2(x)  \\
\end{array}\right)=\left( \begin{array}{c}
\tilde{\psi}_1(x)  \\
\tilde{\psi}_2(x)  \\
\end{array} \right),
\end{equation}
the coupled equation system (\ref{EDSV1}) can be written as
\begin{eqnarray}\label{EDPSDTV}
&&-c\hbar\frac{d}{dx}\left( \begin{array}{c}
\tilde{\psi}_1(x)  \\
\tilde{\psi}_2(x)  \\
\end{array} \right)+V(x)\left( \begin{array}{cc}
-\tau & 0  \\
0  & \tau \\
\end{array} \right)\left( \begin{array}{c}
\tilde{\psi}_1(x)  \\
\tilde{\psi}_2(x)  \\
\end{array} \right)\\ \nonumber
&&=\left( \begin{array}{cc}
\frac{E\ell+mc^2+\zeta \varepsilon}{\tau} &\frac{\varepsilon(1-\ell a
+\tau)-a(1+\tau) mc^2-E(1+\zeta a+\tau)}{\tau(\tau+1)} \\
\frac{\varepsilon(1+\ell b+\tau)-b(1+\tau) mc^2+E(1+\zeta b+\tau)}
{\tau(\tau+1)} &-\frac{E\ell+mc^2+\zeta \varepsilon}{\tau} \\
\end{array} \right)
\left( \begin{array}{c}
\tilde{\psi}_1(x)  \\
\tilde{\psi}_2(x)  \\ \nonumber
\end{array} \right),
\end{eqnarray}
where $a=\zeta+\ell$ and $b=\zeta-\ell$. The coupled equation
system (\ref{EDPSDTV}) can be rewritten in terms of the
first-order differential operators $A$ and $A^{\dag}$, which are
known as ladder operators, in the following form
\begin{eqnarray}
A\tilde{\psi}_1(x)&=&w_1\tilde{\psi}_2(x), \label{APAP1}\\
A^{\dag}\tilde{\psi}_2(x)&=&w_2\tilde{\psi}_1(x),\label{APAP2}
\end{eqnarray}
where the ladder operators are defined as
\begin{eqnarray}
A&=&c\hbar\frac{d}{dx}+\tau V(x)+\frac{E\ell+mc^2+\zeta
\varepsilon}{\tau},\label{esps1}\\
A^{\dag}&=&-c\hbar\frac{d}{dx}+\tau V(x)+\frac{E\ell+mc^2+\zeta
\varepsilon}{\tau},\label{esps2}
\end{eqnarray}
and the eigenvalues associated are
\begin{eqnarray}
w_1&=&\frac{a(1+\tau) mc^2-\varepsilon(1-\ell a+\tau)+ E(1+\zeta
a+\tau)}{\tau(\tau+1)},\label{Defw1}
\\
w_2&=&-\frac{b(1+\tau) mc^2- \varepsilon(1+\ell b+\tau)-E(1+\zeta
b+\tau)}{\tau(\tau+1)}. \label{Defw2}
\end{eqnarray}
For the case in which $\tau \geq 0$, {\it i.e.}
$1+\zeta^2>\ell^2$, the operators $A$ and $A^{\dag}$ are mutually
self-adjoint. On the other hand, for the case in which $\tau$ is
purely imaginary, the operators $A$ and $A^{\dag}$ are no longer
mutually adjoint. In section \ref{Sec5}, we will show that the
last feature is related to the Klein paradox. The equation system (\ref{APAP1})
and (\ref{APAP2}) can be decoupled into the two following Klein-Gordon equations
\begin{eqnarray}
&&H_{-}\tilde{\psi}_1(x)=w\tilde{\psi}_1(x), \label{kge1}\\
&&H_{+}\tilde{\psi}_2(x)=w\tilde{\psi}_2(x), \label{kge2}
\end{eqnarray}
where the effective supersymmetric partners $H_{-}$ and $H_{+}$
are given in terms of the ladder operators as
\begin{eqnarray}
&&H_{-}=A^{\dag}A,\\&&H_{+}=AA^{\dag},
\end{eqnarray}
and the eigenvalue $w$ is written as
\begin{eqnarray}
&&w=w_1w_2 \nonumber\\
&&=\left(\frac{(E^2-m^2c^4) \zeta ^2+(E+\ell mc^2)^2+\varepsilon^2
\left(\ell^2-1\right)+2 \varepsilon (E \ell+mc^2) \zeta
}{\tau^2}\right).\label{espscpsve}
\end{eqnarray}
From the definition of the ladder operators (\ref{esps1}) and
(\ref{esps2}), we observe that the energy $E$ is multiplied by the
parameter $\ell$. This means that the supersymmetric partners
$H_{-}$ and $H_{+}$ are energy dependent. This fact shows
that the presence of electric scalar potentials in the Dirac
equation, {\it i.e.} $\ell \neq 0$, implies that the potentials in the two
Klein-Gordon equations (\ref{kge1}) and (\ref{kge2}) depend on energy.

For the case $\zeta_1=\zeta_3=0$, we have the presence of a
potential purely pseudoscalar in the problem. For this case, it is
necessary to define the matrix $D$ consistently
\begin{equation}
D=\left( \begin{array}{cc}
1 & 1 \\
-1  & 1 \\
\end{array} \right).
\end{equation}
Following a similar procedure as the one developed above, we
obtain for this case the following equations
\begin{eqnarray}
A_{p}\tilde{\psi}_1(x)&=&w_1\tilde{\psi}_2(x),\label{kge3}\\
A_{p}^{\dag}\tilde{\psi}_2(x)&=&w_2\tilde{\psi}_1(x), \label{kge4}
\end{eqnarray}
where the eigenvalues are given by $w_1=E+mc^2$, $w_2=E-mc^2$ and
the ladder operators are
\begin{eqnarray}
A_{p}=c\hbar\frac{d}{dx}+P(x), \\
A_{p}^{\dag}=-c\hbar\frac{d}{dx}+P(x).
\end{eqnarray}
For this case, we observe directly from (\ref{kge3}) and
(\ref{kge4}) that the associated Klein-Gordon equations are not
including an energy-dependent potential.

\section{Factorization method for energy-dependent Hamiltonians}\label{Sec3}

In the previous section, we have shown how an energy-dependent
Hamiltonian associated to the Dirac equation in the presence of
static potentials has been factorized in terms of the ladder
operators. However, until now, in the literature the factorization
method has been only considered for energy-independent
Hamiltonians. For this reason, in this section, we will show that
this method can be consistently generalized for the case of
energy-dependent Hamiltonians. To do it, we will follow an
analogous procedure to the one presented by Dong \cite{DongL} and
Bagchi \cite{Bagchi}. Here, it is important to mention that
several studies about Hamiltonians including energy-dependent
potentials can be found in the literature
\cite{Formanek,Garcia,Hass,Yekken}. To start, we consider two
differential operators $A(E)$ and $A^{\dag}(E)$ mutually
self-adjoint defined by

\begin{eqnarray}
A(E)=c\hbar\frac{d}{dx}+W(x,E), \\
A^{\dag}(E)=-c\hbar\frac{d}{dx}+W(x,E),
\end{eqnarray}
where the real function $W(x,E)$ is generally known as the
superpotential. With the help of this superpotential, the
supersymmetric partners are written as
\begin{eqnarray}
H_{-}(E)=A^{\dag}(E)A(E),\\
H_{+}(E)=A(E)A^{\dag}(E).
\end{eqnarray}
The discrete spectrum of $H_{\pm }(E)$ associated to the
eigenstates $\{\psi^{\pm}_n\}$ is respectively denoted as $\{
E^{\pm}_n \}$, with $n=0,1, \ \ldots $. We restrict now our
interest to the case of unbroken supersymmetry systems, this means
that the ground state eigenvalue of one of these
Hamiltonians is zero. For this kind of systems, there are two
different cases:
\par
(i) If $A_0\tilde{\psi}^{-}_{0}=0$, therefore
\begin{equation}
H_{-}(E_0^-)\tilde{\psi}^{-}_0=A^{\dag}(E_0^-) A(E_0^-)
\tilde{\psi}^{-}_0=0,
\end{equation}
then $\tilde{\psi}^{-}_{0}$ is an eigenstate of $H_{-}$ with a
zero eigenvalue. For $n\geq0$, the relation
\begin{equation}
H_{-}(E_n^-)\tilde{\psi}^{-}_n=A^{\dag}(E_n^-)A(E_n^-)
\tilde{\psi}^{-}_n=E^{-}_n\tilde{\psi}^{-}_n,
\end{equation}
implies
\begin{equation}
H_{+}(E^-_n)(A(E^-_n)\tilde{\psi}^{-}_n)=A(E^-_n)A^{\dag}(E^-_n)A(E^-_n)
\tilde{\psi}^{-}_n=E^{-}_n(A(E^-_n)\tilde{\psi}^{-}_n).
\end{equation}
Thus, for every eigenstate $\tilde{\psi}^-_n$ of $H_{-}$, the
function $A(E^-_n)\tilde{\psi}^{-}_n$ is an eigenstate of $H_{+}$
and its corresponding eigenvalue is $E_n^{-}$. Similarly, we have
\begin{equation}
H_{+}(E^+_n)\tilde{\psi}^{+}_n=A(E^+_n)A^{\dag}(E^+_n)
\tilde{\psi}^{+}_n=E^{+}_n\tilde{\psi}^{+}_n,
\end{equation}
therefore
\begin{equation}
H_{-}(E^+_n)(A^{\dag}(E^+_n)\tilde{\psi}^{+}_n)=A^{\dag}(E^+_n)A(E^+_n)
A^{\dag}(E^+_n)\tilde{\psi}^{+}_n=E^{+}_n(A^{\dag}(E^+_n)\tilde{\psi}^{-}_n).
\end{equation}
Thus, for every eigenstate $\tilde{\psi}^+_n$ of $H_{+}$, the
function $A(E^+_n)\tilde{\psi}^{+}_n$ is an eigenstate of $H_{-}$
and its corresponding eigenvalue is $E_n^{+}$. Consequently, the
following relations are obtained
\begin{eqnarray}
&& E^{+}_n=E^{-}_{n+1}, \ \ \ E^{-}_0=0, \\
&& \tilde{\psi}^{+}_{n}(x)=N^+_n A(E_{n+1}^-)\tilde{\psi}^{-}_{n+1}(x),
\ n=0,\ 1,\ 2\ \ldots, \\
&&
\tilde{\psi}^{-}_{n+1}(x)=N_{n+1}^{-}A^{\dag}(E_{n}^+)\tilde{\psi}^{+}_{n}(x),\
n=1,\ 2\ \ldots,
\end{eqnarray}
where $N^+_n$ and $N^-_n$ are normalization constants.
\par
In order to ensure the conservation of the norm, for the case of
energy-dependent potentials, it has been established that the
probability density should be modified \cite{Formanek}. If the
Hamiltonian for a $n$-eigenstate satisfies the equation
$H_n\tilde{\psi}_n(x)=-\frac{\hbar^2}{2m}\frac{\partial^2
\tilde{\psi}_n(x)}{\partial x^2}+V(x,E_n)\tilde{\psi}_n(x)$, the
wave functions are normalized if the condition
\begin{equation}
\int dx|\tilde{\psi}_n(x)|^2\left[1-\frac{\partial V(x,E_n)
}{\partial E_n}\right]=1
\end{equation}
is satisfied. For the problem considered here, it is not necessary
to modify the probability density because the probability is
conserved by means of the usual normalization condition
\begin{equation}
\int dx(|\psi_1(x)|^2+|\psi_2(x)|^2)=1,
\end{equation}
where $\psi_j(x)=D^{-1}\tilde{\psi}_j(x)$, with $j=1,2$. For the
case of supersymmetric partners, the problem of the dependence on
the energy is entirely due to the method of diagonalization that
is being used. The conservation of probability is ensured because
the external potentials initially considered in the problem are
independent of the energy.
\newline
(ii) If $A^{\dag}(E_0^+)\tilde{\psi}^{+}_{0}=0$, then
$\tilde{\psi}^{+}_{0}$ is an eigenstate of $H_{+}$ with an
eigenvalue zero. For this case, we obtain the following
relations
\begin{eqnarray}
&& E^{-}_n=E^{+}_{n+1}, \ \ \ E^{+}_0=0, \\
&& \tilde{\psi}^{-}_{n}(x)=N_n^-A^{\dag}(E_{n+1}^+)\tilde{\psi}^{+}_{n+1}(x),
\ n=0,\ 1,\ 2\ \ldots, \\
&&
\tilde{\psi}^{+}_{n+1}(x)=N_{n+1}^+A(E_{n}^-)\tilde{\psi}^{-}_{n}(x),\
n=1,\ 2\ \ldots.
\end{eqnarray}

\subsection{Generalization of the shape invariance for energy-dependent Hamiltonians}

As it was mentioned in the introduction, it is possible to obtain
the spectrum algebraically for the case in which the pair of
supersymmetric partners is isospectral and also shape invariant.
But until now, the shape invariance has been studied in the
literature only for energy-independent Hamiltonians. By this
reason, we generalize the shape invariance for a specific class of
energy-dependent Hamiltonians.

Since the supersymmetric partners $H_{+}$ and $H_{-}$ are products
of the operator $A$ and its adjoint $A^{+}$, their eigenvalues are
either zero or positive \cite{Asim}. If these Hamiltonians have
the same dependence on the variable and differ only on other
parameters, {\it i.e.} these are invariant under a discrete
reparametrization (shape invariance), we can obtain the energy
spectrum as well as the eigenfunctions analytically.

For the case in which the ground state of $H_{-}$ has associated a
zero eigenvalue (case (i) of the previous subsection), the shape
invariance condition is given by
\begin{equation}\label{SIcaso1}
H_{+}(a_1,x,E)=H_{-}(a_2,x,E)+R(a_1,E),
\end{equation}
where $a_s=f^{s-1}(a_1)$ ({\it i.e.}, $f$ is applied $s-1$ times). For the particular case in which $R(a_s)$ is not
dependent on $E$, we have
\begin{eqnarray}
E_{1}^-(a_{1})&=&E_{0}^+(a_{1})=E_{0}^{-}(a_{2})+R(a_1)\nonumber\\
&=&R(a_1)
\end{eqnarray}
and
\begin{eqnarray}
E_{2}^-(a_{1})&=&E_{1}^+(a_{1})=E_{1}^{-}(a_{2})+R(a_1)\nonumber\\
&=&E_{0}^{+}(a_{2})+R(a_1)\nonumber\\
&=&E_{0}^{-}(a_{3})+R(a_2)+R(a_1)\nonumber\\  &=&R(a_2)+R(a_1).
\end{eqnarray}
In general, the spectrum is given by
\begin{eqnarray}
E_{n}^-=\sum_{k=1}^{n}R(a_k),
\end{eqnarray}
and the eigenfunction describing the ground state can be written
as
\begin{equation}
\tilde{\psi}^{-}_{0}(x,a_1)=N_0^-\exp^{\left(-\int^{x}
\frac{W(x,E_0^-,a_1)}{c \hbar }dx\right)}.
\end{equation}
From the shape invariant condition Eq. (\ref{SIcaso1}), we have
that $H_{+}(a_1,x,E)$ and $H_{-}(a_2,x,E)$ have the same
eigenfunctions. The eigenfunction describing the ground state
satisfies
\begin{equation}
\tilde{\psi}^+_0(x,a_1)=\tilde{\psi}^-_0(x,a_2)\sim
\exp^{\left(-\int^{x} \frac{W(x,E_0^-,a_2)}{c \hbar }dx\right)},
\end{equation}
therefore
\begin{equation}
\tilde{\psi}^-_1(x,a_1)\sim
A^{\dag}(E_0^+,a_1)\tilde{\psi}^+_0(x,a_1)=A^{\dag}(E_1^-,a_1)\tilde{\psi}^-_0(x,a_2)
\end{equation}
and
\begin{eqnarray}
\tilde{\psi}^-_2(x,a_1)&\sim&
A^{\dag}(E_1^+,a_1)\tilde{\psi}^+_1(x,a_1)=A^{\dag}(E_2^-,a_1)\tilde{\psi}^-_1(x,a_2) \nonumber \\
&=&A^{\dag}(E_2^-,a_1)A^{\dag}(E_1^-,a_2)\tilde{\psi}^-_0(x,a_3).
\end{eqnarray}
For $n>0$, in general, we have
\begin{equation}
\tilde{\psi}^-_n(x,a_1)\sim A^{\dag}(E_n^-,a_1)\dots
A^{\dag}(E_2^-,a_{n-1})A^{\dag}(E_1^-,a_n)\tilde{\psi}^-_0(x,a_{n+1}).
\end{equation}

\subsection{Examples}\label{Example1}

We now consider two examples in which the spectrum of an energy-dependent Hamiltonian
is calculated using the generalizations of the factorization method and shape
invariance previously presented.

\subsubsection{Linear potential}\label{LinPot}

Let us consider the system defined by the $(1+1)$-dimensional
Dirac equation in the presence of a linear potential of the form
$f(x)=\frac{a x}{\tau\zeta_1}$. For this system, the ladder
operators are
\begin{eqnarray}
A&=&c\hbar\frac{d}{dx}+a x+b_n, \label{escV1}\\
A^{\dag}&=&-c\hbar\frac{d}{dx}+a x+b_n, \label{escV2}
\end{eqnarray}
where $b_n=(E_n\ell+mc^2+\zeta \varepsilon)/\tau$. The
supersymmetric partners are
\begin{eqnarray}
H_{-}&=&-(c\hbar)^2\frac{d^2}{dx^2}+a ^2 x^2+2a b_nx+b_n^2-c\hbar
a,\\ H_{+}&=&-(c\hbar)^2\frac{d^2}{dx^2}+a ^2 x^2+2a
b_nx+b_n^2+c\hbar a.
\end{eqnarray}
These Hamiltonians are shape invariants because the condition
(\ref{SIcaso1}) is satisfied
\begin{equation}
H_{+}(a,x)=H_{-}(a,x)+2c\hbar a.\label{sinvcond}
\end{equation}
For this system $a_s=a$, $R(a_s)=2c\hbar a$ and the eigenvalues of
$H_{-}$ are
\begin{equation}
w^{-}_{n}=\sum^{n}_{s=1}R(a_{s})=\sum^{n}_{s=1}2c\hbar a=2c\hbar
an.
\end{equation}
For the case $a>0$, the quantity $w_0$ corresponds to the ground
state eigenvalue and the condition $w_0<w_1<w_2<...$ is satisfied.
For the case $a<0$, we can obtain from (\ref{sinvcond}) the
condition
\begin{equation}
H_{-}(a,x)=H_{+}(a,x)-2c\hbar a,
\end{equation}
which implies that the eigenvalues of $H_{+}$ are
\begin{equation}
w^{+}_{n}=\sum^{n}_{s=1}R(a_{s})=-\sum^{n}_{s=1}2c\hbar a.
\end{equation}
In this form, the eigenvalues of the (1+1)-dimensional Dirac
equation in the presence of static scalar, pseudoscalar and
electric scalar potentials are obtained from (\ref{espscpsve}) and
are given by
\begin{equation} \label{EnEspEx1}
E_n=\frac{-\ell \left(mc^2+\zeta\epsilon \right)\pm \sqrt{\tau^2
\left(\left(\epsilon - mc^2 \zeta \right)^2+2 |a| c \hbar n
\left(1+\zeta ^2\right)\right)}}{1+\zeta ^2}.
\end{equation}
This spectrum is symmetric for $\ell=0$, {\it i.e.} for a
vanishing electric scalar potential $A_t=0$, and is only valid if
the inequality $1+\zeta^2>\ell^2$ is satisfied. The spectrum of a
Dirac particle in ($1+1$) dimensions in the presence of a scalar
field obtained by Long {\it et al}. \cite{Long} is reproduced here
from (\ref{EnEspEx1}), for the case $\zeta=\epsilon=\ell=0$.

\subsubsection{Inversely linear potential}\label{InvPotLin}

Now we consider the system defined by the $(1+1)$-dimensional
Dirac equation in the presence of an inversely linear potential of
the form $f(x)=-\frac{q}{\zeta_1x}$, with $q>0$, where the
variable $x$ is defined in the positive real line
($x\in\mathbb{R}^+$). For this system, the ladder operators are
\begin{eqnarray}
A&=&c\hbar \frac{d}{dx}-\frac{\tilde{\tau}}{x}+f_n,\label{escVC3}\\
A^{\dag}&=&-c\hbar \frac{d}{dx}-\frac{q}{x}+f_n,\label{escVC4}
\end{eqnarray}
with $f_n=(E_n\ell+mc^2+\zeta \varepsilon)/\tau$ and
$\tilde{\tau}=q\tau$. The supersymmetric partners are written as
\begin{eqnarray}
H_{-}&=&-(c\hbar)^2\frac{d^2}{dx^2}+\frac{\tilde{\tau}(\tilde{\tau}-c\hbar)}
{x^2}-\frac{2f_n \tilde{\tau}}{x}+f_n^2,\\
H_{+}&=&-(c\hbar)^2\frac{d^2}{dx^2}+\frac{\tilde{\tau}(\tilde{\tau}+c\hbar)}
{x^2}-\frac{2f_n \tilde{\tau}}{x}+f_n^2.
\end{eqnarray}
It is possible to show that these Hamiltonians are explicitly
shape invariants if the change of variable $\rho=f_n\tilde{\tau}x$ is performed
into the operators (\ref{escVC3}) and (\ref{escVC4}). After this
change of variable, the operators (\ref{escVC3}) and
(\ref{escVC4}) are written respectively as
\begin{eqnarray}
\tilde{A}&=&c\hbar \frac{d}{d\rho}-\frac{\tilde{\tau}}{\rho}+\frac{1}{\tilde{\tau}},
\label{escVC5}\\
\tilde{A}^{\dag}&=&-c\hbar
\frac{d}{d\rho}-\frac{\tilde{\tau}}{\rho}+\frac{1}{\tilde{\tau}}. \label{escVC6}
\end{eqnarray}
These operators are the same that those obtained when the
$(3+1)$-dimensional problem is solved \cite{Coo}. With the new
operators (\ref{escVC5}) and (\ref{escVC6}), the following
equations are satisfied
\begin{eqnarray}\label{hamV}
&&\tilde{H}_{-}\tilde{\psi}^{-}(\rho)=\tilde{A}^{\dag}\tilde{A}\tilde{\psi}^{-}(\rho)=
\tilde{w}\tilde{\psi}^{-}(\rho)= \frac{w}{f_n^2\tilde{\tau}^2}\tilde{\psi}^{-}(\rho),\\
&&\tilde{H}_{+}\tilde{\psi}^{+}(\rho)=\tilde{A}\tilde{A}^{\dag}\tilde{\psi}^{+}(\rho)=
\tilde{w}\tilde{\psi}^{+}(\rho)=
\frac{w}{f_n^2\tilde{\tau}^2}\tilde{\psi}^{+}(\rho).
\end{eqnarray}
For this case, the shape invariance leads to
\begin{equation}
\tilde{H}_{+}(\tilde{\tau},\rho)=\tilde{H}_{-}(\tilde{\tau}+c\hbar,\rho)+
\frac{1}{\tilde{\tau}^2}-
\frac{1}{(\tilde{\tau}+c\hbar)^2},
\end{equation}
\begin{equation}
a_2=\tilde{\tau}+c\hbar, \ \ \ \ \ a_1=\tilde{\tau}, \ \ \ \ \
R(a_2)=\frac{1}{\tilde{\tau}^2}-\frac{1}{(\tilde{\tau}+c\hbar)^2}.
\end{equation}
Thus, the eigenvalues for this case are
\begin{equation}
\frac{w_n}{f_n^2\tilde{\tau}^2}=\sum^{n}_{j=1}
R(a_j)=\frac{1}{\tilde{\tau}^2}-\frac{1}{(\tilde{\tau}+nc\hbar)^2},
\end{equation}
implying that the spectrum is given by the solutions of
\begin{equation}
w_n=f_n^2\left(1-\frac{\tilde{\tau}^2}{(\tilde{\tau}+nc\hbar)^2}\right).
\end{equation}
In general, the expression for the spectrum is complicated but we
can obtain simple analytical expressions of this spectrum for two
special cases:

\textit{(i)} For $l=0$, we have the case of a mix between
scalar and pseudoscalar potentials and there is no presence of
an electric scalar potential. For this case, the spectrum is given
by
\begin{equation} \label{EnEspEx2Case1}
E_n=\pm
\frac{(mc^2+\zeta\varepsilon)}{\tau}\sqrt{\frac{1}{\tau^2}- \frac{
q^2}{(q\tau+nc\hbar)^2}
+\frac{(mc^2\zeta-\varepsilon)^2}{(mc^2+\varepsilon\zeta)^2}},
\end{equation}
where $\tau=\sqrt{1+\zeta^2}$. If $\zeta=0$ and $\varepsilon=0$,
the energy spectrum that we obtain from (\ref{EnEspEx2Case1})
corresponds to the one presented by Castro \cite{Castro2}, for the
case of a symmetric scalar potential of the form
$V(x)=-\frac{k}{|x|}$.

\textit{(ii)} For $\zeta=0$ and $\varepsilon=0$, we have the
case of a mix between scalar and electric scalar potentials. For
this case, the spectrum is given by
\begin{equation} \label{EnEspEx2Case2}
E_{n}=
\frac{mc^2}{(ql)^2+(nc\hbar+\tilde{\tau})^2}\left(-q^2l\pm(nc\hbar+
\tilde{\tau})\sqrt{(nc\hbar+\tilde{\tau})^2-\tilde{\tau}^2}\right),
\end{equation}
where $\tilde{\tau}=q\sqrt{1-l^2}$. This result is in agreement
with the spectrum found by Castro \cite{Castro2005}, who has
shown that the problem can be mapped into a Sturm-Liouville
problem in such a way that the spectrum can be obtained. For the case of
two and three spacial dimensions, Xing {\it et al.} have solved
the Dirac equation in the presence of scalar and electric scalar
potentials \cite{Xing2008}, by using the same method that we have
developed in this work.

\section{The (2+1)-dimensional Dirac equation}\label{Sec4}

The (2+1)-dimensional Dirac equation in the presence of scalar
$V(x,y)$ and gauge $A_{\mu}(x,y)$ potentials is given by
\begin{equation}\label{2DDE1}
\left[i\gamma^{\mu}\left(\hbar\partial_{\mu}+i\frac{e}{c}A_{\mu}(x,y)\right)-\frac{1}{c}V(x,y)-mc\right]\Psi(\vec{r},t)=0,
\end{equation}
where the Dirac matrices $\gamma^{\mu}$, with $\mu=0,1,2$, are the
generators of the Clifford algebra in the three-dimensional flat
spacetime
\begin{equation}
\{\gamma^{\mu},\gamma^{\nu}\}=2\eta^{\mu \nu}.
\end{equation}
There are two non-equivalent representations of the Dirac matrices
in the (2+1) dimensional case
\begin{equation}
\gamma^{0}=\sigma^3=\left( \begin{array}{cc}
1 & 0 \\
0 & -1 \\
\end{array} \right), \ \ \  \gamma^{1}=si\sigma^1=s\left( \begin{array}{cc}
0 & i \\
i & 0 \\
\end{array} \right), \ \ \  \gamma^{2}=i\sigma^2=\left( \begin{array}{cc}
0 & 1 \\
-1 & 0 \\
\end{array} \right),
\end{equation}
where the parameter $s=\pm1$ characterizes the two possible values
of the electron spin projection and $\sigma^{i}$ are the usual
Pauli matrices. Taking into account that
$\beta=\gamma^{0}=\sigma^3$,
$\alpha_1=\gamma^{0}\gamma^{1}=-s\sigma^2$ and
$\alpha_2=\gamma^{0}\gamma^{2}=\sigma^1$, the Dirac equation
(\ref{2DDE1}) can be rewritten as follows
\begin{equation} \label{2DDE2}
i\hbar\frac{\partial}{\partial
t}\Psi(x,y,t)=\mathcal{H}\Psi(x,y,t)=c\alpha_1 p_x+c\alpha_2
p_y+\beta mc^2+\mathcal{V},
\end{equation}
where
\begin{equation}
\mathcal{V}=I eA_t(x,y) + \sigma^1 eA_y(x,y)-s\sigma^2
eA_x(x,y)+\sigma^3V(x,y),
\end{equation}
with $I$ representing the $2×2$ identity matrix. The potential
matrix $\mathcal{V}$ corresponds to the most general combination
of Lorentz structures because there are only four linearly
independent $2×2$ matrices \cite{Castro4}. If we write the spinor
$\Psi(x,y,t)$ as
\begin{equation}
\Psi(x,y,t)=e^{-\frac{iEt}{\hbar}}\left( \begin{array}{c}
\psi^{(1)}(x,y)  \\
\psi^{(2)}(x,y) \\
\end{array} \right),
\end{equation}
then the Dirac equation (\ref{2DDE2}) can be written as a coupled
equation system
\begin{equation}\label{2DDE3}
-sc\hbar\frac{\partial \psi^{(1)}}{\partial
x}-ic\hbar\frac{\partial \psi^{(1)}}{\partial y} +e(A_y-isA_x)
\psi^{(1)}-(E+mc^2+ V-eA_t)\psi^{(2)}=0
\end{equation}
\begin{equation} \label{2DDE4}
-sc\hbar\frac{\partial \psi^{(2)}}{\partial
x}+ic\hbar\frac{\partial \psi^{(2)}}{\partial y}
-e(A_y+isA_x)\psi^{(2)}+(E - mc^2 - V - eA_t)\psi^{(1)} =0.
\end{equation}
With the purpose to study a set of soluble systems using the
factorization method, we will restrict the following treatment to
the case in which the potentials depend only on one spatial
coordinate, that we take as the coordinate $x$. With this
restriction, the wavefunctions $\psi^{(1,2)}(x,y)$ can be written
in the form
\begin{equation}
\psi^{(1,2)}(x,y)=e^{\frac{iky}{\hbar}}\psi_{1,2}(x),
\end{equation}
and then the coupled equation system (\ref{2DDE3}) and
(\ref{2DDE4}) is written as
\begin{eqnarray}\label{SPMG1} \nonumber
-sc\hbar\frac{d}{dx}\left( \begin{array}{c}
\psi_1(x)  \\
\psi_2(x)  \\
\end{array} \right)+\left( \begin{array}{cc}
-iesA_x(x)+eA_y(x) & eA_t(x)-V(x)\\
-eA_t(x)-V(x) & -iesA_x(x)-eA_y(x) \\
\end{array} \right)\left( \begin{array}{c}
\psi_1(x)  \\
\psi_2(x)  \\
\end{array} \right)&&\\
=\left( \begin{array}{cc}
-c k & E+mc^2 \\
-E+mc^2 & c k \\
\end{array} \right)\left( \begin{array}{c}
\psi_1(x)  \\
\psi_2(x)  \\
\end{array} \right).&&
\end{eqnarray}
We can observe that the equation system (\ref{SPMG1}) has the same
form as the equation system (\ref{EDCS}) of the
$(1+1)$-dimensional case, in such a way that the potentials
$V(x)$, $eA_y(x)$, $A_x(x)$ and $A_t(x)$ of the
$(2+1)$-dimensional case play the same role respectively as the
potentials $P(x)$, $-V(x)$, $-A_x(x)$ and $-A_t(x)$ of the
$(1+1)$-dimensional case. In general the $A_x$ component
can be removed by a gauge transformation, without loss of
generality, we impose $A_x=0$ in the following treatment.

\subsection{Mix of scalar, electric scalar and vectorial potentials}

The mix of vectorial, scalar and electric scalar potentials
depending on the $x$-variable for the (2+1) dimensional case is
analogous to the one discussed in section (\ref{Mispv}) for the
(1+1) dimensional case. As it was performed in the (1+1)
dimensional case, here we also impose the same functional form
$g(x)$ for all the potentials which are present in the problem. In
this form, we write $eA_y(x)=\lambda_1 g(x)$, $V(x)=\lambda_2g(x)$
and $eA_t(x)=\lambda_3g(x)$, where $\lambda_1$, $\lambda_2$ and
$\lambda_3$ are constants. Following a similar procedure as the
one performed for the (1+1)-dimensional case, in section
(\ref{Mispv}), we obtain that
\begin{eqnarray}\label{EDPSEM2}
&&-cs\hbar\frac{d}{dx}\left( \begin{array}{c}
\tilde{\psi}_1(x)  \\
\tilde{\psi}_2(x)  \\
\end{array} \right)+eA_y(x)\left( \begin{array}{cc}
\tau & 0  \\
0  & -\tau \\
\end{array} \right)\left( \begin{array}{c}
\tilde{\psi}_1(x)  \\
\tilde{\psi}_2(x)  \\
\end{array} \right)\\ \nonumber
&&=\left( \begin{array}{cc}
-\frac{E_n\nu+\lambda mc^2+ck}{\tau} &\frac{\lambda(mc^2-ck\lambda)
+E\nu+ck\nu^2 +(E\lambda+mc^2\nu)\tau }{\tau(1+\nu)} \\
\frac{\lambda(mc^2-ck\lambda)+E_n\nu+ck\nu^2 -(E\lambda+mc^2\nu)
\tau }{\tau(1-\nu)} &\frac{E\nu+\lambda mc^2+ck}{\tau} \\
\end{array} \right)
\left( \begin{array}{c}
\tilde{\psi}_1(x)  \\
\tilde{\psi}_2(x)  \\
\end{array} \right),
\end{eqnarray}
where $\lambda=\lambda_2/\lambda_1$, $\nu=\lambda_3/\lambda_1$ and
$\tau=\sqrt{1+\lambda^2-\nu^2}$. For this case, the ladder
operators are
\begin{eqnarray}
A&=&sc\hbar\frac{d}{dx}-\tau eA_y(x)-\frac{E\nu+\lambda mc^2+ck}{\tau}, \\
A^{\dag}&=&-sc\hbar\frac{d}{dx}-\tau eA_y(x)-\frac{E\nu+\lambda mc^2+ck}{\tau}.
\end{eqnarray}
If the condition $1+\lambda ^2>\nu^2$ is satisfied, the symmetric
partner Hamiltonians are energy dependent and the operators $A$
and $A^{\dag}$ are mutually self-adjoint. For the other case,
$\tau$ is purely imaginary and the operators $A$ and $A^{\dag}$
are no longer mutually self-adjoint. The supersymmetric partners
for this case are
\begin{eqnarray}
&&H_{-}\tilde{\psi}_1(x)=A^{\dag}A\tilde{\psi}_1(x)=\varpi\tilde{\psi}_1(x), \\
&&H_{+}\tilde{\psi}_2(x)=AA^{\dag}\tilde{\psi}_2(x)=\varpi\tilde{\psi}_2(x),
\end{eqnarray}
where
\begin{eqnarray}\label{espmagelec}
\varpi=\left(\frac{E^2(1+\lambda^2)-(mc^2-ck\lambda)^2+2E(ck+\lambda mc^2)
\nu+(c^2k^2+m^2c^4)\nu^2}{\tau^2}\right).
\end{eqnarray}

\subsubsection{Mix of a scalar linear potential and a constant electromagnetic field}

Now we will solve the Dirac equation in the presence of a scalar
linear potential $V=Cx$, a magnetic field perpendicular to the
$xy$-plane with a negative direction over the $z$ axis
$eA_y(x)=-e\mathbb{B}x$ and a constant electric field in the $x$
direction $A_t=-\mathbb{E}x$. For this case
$\lambda=-C/e\mathbb{B}$, $\nu=\mathbb{E}/\mathbb{B}$ and
$\tau=\sqrt{1+(C/e\mathbb{B})^2-(\mathbb{E}/\mathbb{B})^2}$ and
the ladder operators for the case $s=1$ are defined as
\begin{eqnarray}\label{escV}
A=\frac{d}{dx}+\tau e\mathbb{B}x-\frac{E\nu+\lambda
mc^2+ck}{\tau}, \label{escV1}\\ A^{\dag}=-\frac{d}{dx}+\tau
e\mathbb{B} x-\frac{E\nu+\lambda mc^2+ck}{\tau}.\label{escV2}
\end{eqnarray}
The definition of the ladder operators changes for the case
$s=-1$, in such a way that $A\leftrightarrow A^{\dag}$. The
problem that we are considering here is completely analogue to the
one studied in the (2+1)-dimensional case, which was presented in
section (\ref{LinPot}). If we follow the same procedure as the one
developed in the (1+1)-dimensional case, the eigenvalues are
obtained from the solution of Eq. (\ref{espmagelec}).
These eigenvalues are given by
\begin{equation}
E_n=\frac{-\nu \left(\lambda mc^2+ck\right)\pm \sqrt{\tau^2
\left(\left(ck\lambda - mc^2 \right)^2+2\tau|e\mathbb{B}|c
\hbar n \left(1+\lambda ^2\right)\right)}}{1+\lambda ^2}.
\end{equation}
For the case in which $m=\lambda=0$, the energy spectrum that we
obtain is the same as the one found by Lukose {\it et al.}
\cite{Lukose}. These authors have used the fact that if
$\mathbb{B}>\mathbb{E}$, then it is possible always to boost to a
frame of reference where the electric field vanishes and the
magnetic field is reduced. In this form, these authors have
obtained the spectrum associated to this new magnetic field and
finally they apply the inverse boost transformation to obtain the
spectrum \cite{Lukose}. We note that the range of validity of the
procedure developed by these authors is equivalent to the one
developed here ($\mathbb{B}>\mathbb{E}$). Similar
conditions for the existence of bound states for the
(3+1)-dimensional Klein-Gordon and Dirac equations in the presence
of a mix between electric and magnetic fields were found by Adame
{\it et al.} \cite{Adame1992}. Other possible application of the
method presented in this work, related to the most general
combination of potentials having the same functional form, is the
obtention of the spectrum for the system defined by the
($2+1$)-dimensional Dirac oscillator in the presence of an
external uniform magnetic field \cite{Quimbay2013}.
\section{Supersymmetric partners and Klein paradox}\label{Sec5}

The $(1+1)$-dimensional Schrödinger equation in the presence of a
linear scalar potential of the form $V(x)=l+k|x|$ has been
satisfactorily solved and leads to the existence of bound states
\cite{Vall}. However, for the $(1+1)$-dimensional Dirac equation
in the presence of a linear electric scalar potential of the form
$A_t=kx$, it has been found that there is no bound state
solution \cite{Galic,Sara}. A simple form to obtain this result is
to set the gauge potential as $A_{\mu}=(A_t(x),0)=(kx,0)$ in the
Eq. (\ref{ED}). For this case, we can obtain from
(\ref{DEUNOMASUNO}) the following decoupled equation system
\begin{eqnarray}\nonumber
H_{\mp}\psi_{1(2)}(x)&=&\left[-c^2\hbar^2\frac{d^2}{dx^2}-e^2k^2x^2-2ekxE-
E^2\pm i e\hbar k \right]\psi_{1(2)}(x)\\ \label{Asinto}
&=&-m^2c^4\psi_{1(2)}(x),
\end{eqnarray}
where $H_{\mp}$ are not supersymmetric partners because these
Hamiltonians cannot be constructed in terms of the two operators
$A$ and $A^{\dag}$ mutually self-adjoints. We observe that in
Eq. (\ref{Asinto}) the term $e^2k^2x^2$ is dominant and the asymptotic
behavior of the eigenfunctions are determined by the equation
\begin{equation}\label{Asinto1}
\left[c^2\hbar^2\frac{d^2}{dx^2}+e^2k^2x^2\right] {\psi}_{1(2)}
(|x|\rightarrow \infty)=0.
\end{equation}
The behavior of Eq. (\ref{Asinto1}) has been
very well studied by Saradzhev \cite{Sara}, who has show that
the solutions do not belong to $L^2$, thus there are not bound
states. The absence of bound state solutions in this problem has
been attributed to the Klein paradox \cite{Castro2,Galic}.
Additionally, the problem of a mixed scalar-electric scalar linear
potentials in ($1+1$) dimensions has been studied by Castro
\cite{Castro4}, who found bound state solutions for the case in
which the scalar coupling has the sufficient intensity compared to
the electric scalar coupling. The last result is a particular case
of the problem that we have considered in the example
\ref{LinPot}.
\par
Additionally to the above example, it is possible to find in the
literature more examples in which if the intensity of the electric
scalar potential is higher than the one of the scalar potential
then there are not bound state solutions. For instance, although
the ($1+1$)-dimensional Schrödinger equation in the presence of an
inverse linear potential has been rightly solved by Ran {\it et
al.} \cite{Ran}, for the $(1+1)$-dimensional Dirac equation in the
presence of an electric scalar inverse linear potential of the
form $A_t(x)=-k/|x|$, it has been observed the non-existence of
bound states solutions \cite{Shi1}. As it was noted previously in
the example \ref{InvPotLin}, for the case in which the coupling
intensity of the electric scalar potential does not exceed the coupling
intensity of the scalar and pseudoscalar potentials (this means
for the case in which $\tau$ is real), the ($1+1$)-dimensional
Dirac equation in the presence of a mix between scalar,
pseudoscalar and electric scalar inverse linear potentials has
associated bound state solutions. This result is in agreement with
the condition found by Castro, for the case of mixed scalar-electric
scalar potentials \cite{Castro2005}.
\par
The solution of the (1+1)-dimensional Dirac equation in the
presence of a scalar potential $V(x)$ leads to energy levels $E$
which are symmetric with respect to $E=0$ (see, for instance,
references \cite{Castro2,Nogami,Couti}). This symmetric behavior
can be understood taking into account that if $\Psi(x,t)$
represents an eigenfunction of Eq. (\ref{ED}), for the
case $A_t(x)=0$ and $V(x)\neq0$, having an associated energy level
$+E$, it is possible to prove that $\sigma^3\Psi^*(x,t)$
represents also an eigenfunction of the same equation but having
an associated energy level $-E$. However, if now $\Psi(x,t)$ is an
eigenfunction of Eq. (\ref{ED}), but for the case
$V(x)=0$ and $A_t(x)\neq0$, having an associated energy level
$+E$, then $\sigma^3\Psi^*(x,t)$ is not an eigenfunction of the
same equation having an associated energy level $-E$. Therefore,
the solution of the Dirac equation in the presence of an electric
scalar potential $A_t(x)$ leads to energy levels which are not
symmetric with respect to $E=0$. This is an example that shows one
of the differences between the physics which is described by the
Dirac equation in the presence of a scalar potential $V(x)$ and an
electric scalar potential $A_t(x)$, considering that both
potentials have the same functional form.
\par
We have studied in sections \ref{Sec2} and \ref{Sec4} the Dirac
equation in $(1+1)$ and $(2+1)$ dimensions in the presence of
static potentials having the same functional form. We have found
that the Dirac equations (\ref{EDPSDTV}) and (\ref{EDPSEM2}) can
be looked like Dirac equations in the presence of scalar
potentials with rescaled couplings $V(x)\rightarrow \tau V(x)$.
This means that if the coupling of the electric scalar potential
$A_t$ has an intensity for which the parameter $\tau$ is real,
then for this case there is no mix between the positive and
negative energy states, in the same way as the scalar potential
does not. The condition for $\tau$ to be real implies that if we
can build two Schrödinger supersymmetric partner Hamiltonians for
the Dirac equation, then the Dirac sea is stable and the
variations of the couplings in the potentials provides a way to
avoid the Klein paradox.
\par
It is worth noting that the condition for $\tau$ is necessary but
not sufficient for establishing the absence of the Klein
paradox. To illustrate the validity of this sentence, we consider
the ($1+1$)-dimensional Dirac equation in the presence of mixed
scalar-electric scalar step potential. For this case, the sufficient
condition for the spontaneous production of particle-antiparticle
pairs is $l>1+2mc^2/V$, where $V$ is the scalar potential. In our case,
if $\tau$ is not real, means that $l>1$. Therefore, for the
interval $1+2mc^2/V>l>1$ there is not the Klein paradox, but our condition
cannot predict the fact that the vacuum is stable
for this interval. In this sense, this condition is necessary but
is not sufficient.

\section{Conclusions}\label{Sec6}

In this work we have studied the ($1+1$) and ($2+1$)-dimensional
Dirac equations in the presence of static external potentials. We
have considered the mix of the potentials as the most general
combination of Lorentz structures, for the case in which the
potentials have the same functional. After we have diagonalized
this problem, we have arrived to rewrite the Dirac equation in
terms of two first-order differential operators, which are called
ladder operators, and we have showed that the problem has been
mapped to a Dirac equation in the presence of a scalar potential
with rescaled coupling. If the ladder operators $A$ and $A^{\dag}$
are mutually self-adjoint, then the Dirac sea is stable,
therefore, there is not the Klein paradox and the Dirac particles
can be confined. The factorization method has been also
generalized for energy-dependent Hamiltonians and the shape
invariance for a restricted class of energy-dependent
Hamiltonians.

\section*{Acknowledgments}
J. A. Sánchez thanks CAPES for financial support.



\begin{thebibliography}{999}
\bibitem{Gerri1} R. Gerritsma, G. Kirchmair, F. Zähringer, E. Solano, R. Blatt, and C. F. Roos, Nature (London) 463 (2010) 68.

\bibitem{Gerri2} R. Gerritsma, B. P. Lanyon, G. Kirchmair, F. Z\"{a}hringer, C.
Hempel, J. Casanova, J. J. Garcia-Ripoll, E. Solano, R. Blatt, and
C. F. Roos, Phys. Rev. Lett. 106 (2011) 060503.

\bibitem{Casanova} J. Casanova, J. J. García-Ripoll, R. Gerritsma, C. F. Roos, and E. Solano,
Phys. Rev. A 82 (2010) 020101.

\bibitem{Castro2006} N. M. R. Peres, F. Guinea, and A. H. Castro Neto, Phys. Rev. B 73 (2006) 205408.

\bibitem{Kats} M. I. Katsnelson, K. S. Novoselov, and A. K. Geim,  Nature Phys. 2 (2006) 620.

\bibitem{BagGit} V. B. Bagrov and D. M. Gitman, \textit{Exact solutions of relativistic wave equations},
Kluwer Academic Publishers, 1990.

\bibitem{Thal} B. Thaller,\textit{The Dirac equation},
Springer-Verlag Berlin Heildeberg, 1992.

\bibitem{Coo} F. Cooper, A. Khare, and U. Sukhatme, Phys. Rep. 251 (1995) 267.

\bibitem{Adame1990} F. Domínguez-Adame and M. A. González, Europhys. Lett. 13 (3) (1990) 193.

\bibitem{Adame1992} F. Dominguez-Adame and B. Méndez, Nuovo Cimento B 107 (1992) 489.

\bibitem{Long} C. Y. Long and C. Y. Qin, Chin. Phys. 16 (2007) 897.

\bibitem{Castro4} A. S. de Castro, Phys. Lett. A 305 (2002) 100.

\bibitem{Castro2003a} A. S. de Castro and W. G. Pereira, Phys. Lett. A 308 (2003) 131.

\bibitem{Castro2003b} A. S. de Castro, Phys. Lett. A 318 (2003) 40.

\bibitem{Castro2} A. S. de Castro, Phys. Lett. A 328 (2004) 289.

\bibitem{Castro2005} A. S. de Castro, Ann. Phys. 316 (2005) 414.

\bibitem{2007} L. B. Castro, A. S. de Castro and M. Hott, Int. J. Mod. Phys. E, 16 (2007) 3002.

\bibitem{Shi1} S. H. Dong, J. Phys. A 36 (2003) 4977.

\bibitem{Shi2} S. H. Dong and Z. Q. Ma, Phys. Lett. A 312 (2003) 78.

\bibitem{Zarr} S. Zarrinkamar, A. Rajabi, and H. Hassanabadi, Ann. Phys. 325 (2010) 1720.

\bibitem{Hughes} R. Hughes, V. A. Kostelecky and M. M. Nieto, Phys. Rev. D 34 (1986) 1100.

\bibitem{Beckers} J. Beckers and N. Debergh, Phys. Rev. D 42 (1990) 1255.

\bibitem{bookdirac} P. A. M. Dirac, \textit{The Principles of Quantum Mechanics} (Clarendon Press, Oxford) 1935.

\bibitem{Schro1} E. Schrodinger, Proc. R. Irish Acad. A 46 (1940) 9; 46 (1940) 183; 47 (1941) 53.

\bibitem{Infeld} L. Infeld and T. E. Hull, Rev. Mod. Phys. 23(1) (1951) 21-68.

\bibitem{Darboux1} G. Darboux, C. R. Acad, Sci. (Paris) 94 (1882) 1456.

\bibitem{Witten} E. Witten, Nucl. Phys. B 188 (1981) 513.

\bibitem{Gende} L. Gendenshtein, JETP Lett. 38 (1983) 356.

\bibitem{MartinezR} R. P. Martínez y Romero, M. Moreno, and A. Zentella, Phys. Rev. D 43 (6) (1991) 2036.

\bibitem{Benitez} J. Benitez, R. P. Martinez y Romero, H. N. Nuñez-Yépez and
A. L. Salas-Brito, Phys. Rev. Lett. 64 (1990) 1643.

\bibitem{Sinha1} A. Sinha and P. Roy, Mod Phys. Lett. A, 20 (31) (2005) 2377.

\bibitem{Sinha2} A. Sinha and P. Roy, Int. J. Mod. Phys A, 21 (2006) 5807.

\bibitem{Santos1} V. G. dos Santos, A. de Souza Dutra, and M. B. Hott, Phys. Lett. A, 373 (38) (2009) 3401.

\bibitem{DongL} S. H. Dong, \textit{Factorization Method in Quantum Mechanics}, Springer, 2007.

\bibitem{Bagchi} B. K. Bagchi, \textit{Supersymmetry in quantum and classical mechanics}. Chapman and Hall-CRC, 2010.

\bibitem{Formanek} J. Formánek, R. J. Lombard, and J. Mare\u{s}, Czech. J. Phys. 54(3) (2004) 289.

\bibitem{Garcia} J. García-Martínez, J. García-Ravelo, J. J. Pena, and A. Schulze-Halberg, Phys. Lett. A 1 (2009) 5.

\bibitem{Hass} H. Hassanabadi, E. Maghsoodi, R. Oudi, S. Zarrinkamar, H. Rahimov, Eur. Phys. J. Plus 127 (2012) 120.

\bibitem{Yekken} R. Yekken, M. Lassaut, and R. J. Lombard, Few-Body Systems (2013) 1.

\bibitem{Asim}  A. Gangopadhyaya, J. V. Mallow, and C. Rasinariu, \textit{Supersymmetric Quantum Mechanics: An Introduction} World
    Scientific, (2010).

\bibitem{Xing2008} J. G. Xing and R. Z. Zhou, Commun. Theor. Phys. 49 (2008) 319.

\bibitem{Lukose} V. Lukose, R. Shankar, and G. Baskaran, Phys. Rev. Lett. 98 (2007) 116802.

\bibitem{Quimbay2013} C. Quimbay and P. Strange, (2013). arXiv:1312.5251.

\bibitem{Vall} O. Vallée and M. Soares M, \textit{Airy Functions and Applications to Physics}, London:
Imperial College Press, 2004.

\bibitem{Galic} H. Galic, Am. J. Phys. 56 (1988) 312.

\bibitem{Sara} F. Saradzhev, J. Phys. A: Math. Gen. 34 (2001) 1771.

\bibitem{Ran} Y. Ran, L. Xue, S. Hu, and R. K. Su, J. Phys. A 33 (2000) 9265.

\bibitem{Nogami} F. A. B. Coutinho abd Y. Nogami, Phys. Lett. A 124 (1987) 211.

\bibitem{Couti} F. A. B. Coutinho, Y. Nogami, and F. M. Toyama, Am. J. Phys. 56 (10) (1988) 904.

\end{thebibliography}
\end{document}